\documentclass[twocolumn, twocolappendix, astrosymb, times]{aastex631}

\usepackage{graphicx}	

\newcommand{\pluto}{\textsc{pluto }}
\newcommand{\cloudy}{\textsc{cloudy }}
\newcommand{\ionr}[2]{#1 \textsc{#2}}

\newcommand{\psec}{\,\mbox{ s}^{-1} }
\newcommand{\cc}{\,\mbox{ cm}^{3} }
\newcommand{\pcc}{\,\mbox{ cm}^{-3} }
\newcommand{\pcm}{\,\mbox{ cm}^{-1} }
\newcommand{\pcmsq}{\,\mbox{ cm}^{-2} }
\newcommand{\mpcc}{\, m_{\rm p} \pcc}
\newcommand{\msun}{\,\mbox{ M}_{\odot}}
\newcommand{\mpy}{\,\mbox{ M}_{\odot} \mbox{yr}^{-1} }
\newcommand{\ergps}{\,\mbox{ erg s}^{-1} }
\newcommand{\ergpspcmsq}{\,\mbox{ erg s}^{-1} \mbox{cm}^{-2} }

\newcommand{\photps}{\,\mbox{photons s}^{-1} }
\newcommand{\photpskeV}{\,\mbox{photons s}^{-1} \mbox{ keV}^{-1} }
\newcommand{\kmps}{\,\mbox{km s}^{-1}}
\newcommand{\tcool}{\,t_{\rm cool} }
\newcommand{\tdyn}{\,t_{\rm dyn} }

\newcommand{\nei}{\textit{NEI} }

\newcommand{\oi}{\ionr{O}{i} }
\newcommand{\oii}{\ionr{O}{ii} }
\newcommand{\oiii}{\ionr{O}{iii} }
\newcommand{\oiv}{\ionr{O}{iv} }
\newcommand{\ovi}{\ionr{O}{vi} }
\newcommand{\ovii}{\ionr{O}{vii} }
\newcommand{\oviii}{\ionr{O}{viii} }

\newcommand{\ddr}{\frac{\partial}{\partial r}}
\newcommand{\ddt}{\frac{\partial}{\partial t}}
\newcommand{\angs}{\mbox{\normalfont\AA}}
\newcommand{\pie}{\textit{PIE} }
\newcommand{\piesr}{\textit{PIE-SR} }
\newcommand{\neisr}{\textit{NEI-SR} }

\shorttitle{self-ionizing galactic winds}
\shortauthors{Sarkar, Sternberg \& Gnat}
\submitjournal{ApJ}

\begin{document}
\title{Self-ionizing galactic winds} 
\correspondingauthor{Kartick C. Sarkar}
\email{sarkar.kartick@mail.huji.ac.il, kartick.c.sarkar100@gmail.com}

\author[0000-0002-7767-8472]{Kartick C. Sarkar}
\affiliation{Racah Institute of Physics, The Hebrew University of Jerusalem, 91904, Israel}

\author{Amiel Sternberg}
\affiliation{School of Physics and Astronomy, Tel Aviv University, Ramat Aviv, 69978, Israel}
\affiliation{Center for Computational Astrophysics, Flatiron Institute, 162 5th Avenue, 10010, New York, NY, USA}
\affiliation{Max-Planck-Institut fur Extraterrestrische Physik (MPE), Giessenbachstr., 85748, Garching, FRG}

\author{Orly Gnat}
\affiliation{Racah Institute of Physics, The Hebrew University of Jerusalem, 91904, Israel}

\begin{abstract}
We present hydrodynamical simulations of galactic winds from star-forming galaxies including
non-equilibrium ionization and frequency-dependent radiative transfer, processes that have remained largely unaccounted for in galactic wind studies. We consider radiation from massive stars, the metagalactic UV/X-ray background, and the self-radiation of the 
supernovae heated gas. We compare our results to classical galactic wind solutions and show the importance of our newly included physical processes toward observations of ions such as \ionr{O}{iii}, \ionr{O}{vi}, \ionr{O}{vii} and \ionr{O}{viii} plus the observable soft X-ray spectra. Non-equilibrium ionization is reflected in over-ionized gas compared to equilibrium solutions, leading to much enhanced column densities of highly ionized species.
The wind produces excess soft X-ray ($E\gtrsim 100$ eV) radiation that is
several orders of magnitude higher compared to the metagalactic background. 
This radiation ionizes the higher ions (such as \ionr{O}{vii}) somewhat, but affects the lower ions (such as \ionr{O}{iii}) significantly.
We predict that the observable X-ray spectra should contain the signatures of such non-equilibrium effects, especially in X-ray lines such as \ionr{O}{vii} and \ionr{O}{viii}. Simple estimates suggest that both the temperature and density of the winds may be overestimated by factors of a few to almost 2 orders of magnitude using simple equilibrium models. We conclude that both the non-equilibrium ionization and the radiation from the wind itself need to be considered for proper modeling of the optical/UV/X-ray emitting plasma in galactic winds.
\end{abstract}


\section{Introduction}
\label{sec:intro}
Galactic winds form a crucial part of the baryonic cycle in galaxies as they can affect the star formation in the interstellar medium \citep{Larson1974, Dekel1986, Kim2015}, and transport mass, energy and heavy metals to the circumgalactic and intergalactic medium \citep{Songaila1996, Nath1997, Borthakur2013, Li2013a, Sarkar2015a, Suresh2015a, Fielding2017, Faerman2020}. Studies of galactic winds and their dependence on star formation rate (SFR), and galaxy interstellar and circumgalactic medium (ISM and CGM) properties are critical for understanding the evolution of galaxies. 

Theoretically, the strength of feedback in a galaxy is often quantified in terms of the mass and energy outflow rates associated with star-formation and/or active nuclei. Numerical simulations of ISM patches \citep{Girichidis2016b, Kim2018} or global outflows in galaxies \citep{Muratov2015, Sarkar2015a, Fielding2017, Nelson2019, Schneider2020, Li2020, Pandya2021} suggest that the outflows are multiphased. The mass carried out in the hot phase ($T\gtrsim 10^5$ K) in such outflows is found to be comparable to the warm phase ($T\lesssim 10^5$ K). However, the energy is predominantly carried out by the hot phase. It is, therefore, crucial to measure both the hot and warm phases through observational lenses to learn more about the mass and energy outflows.

Observationally, the outflow rates of mass/metals/energy from starburst galaxies are estimated either from absorption signatures \citep{Heckman1990, Heckman2000, Martin2005, Rupke2005, Steidel2010, Chisholm2016, Chisholm2017} or from emission signatures \citep{Lynds1963, Forgarty2012, Heckman2015, Hayes2016, Rupke2019, Schreiber2019} of lines such as, H$\alpha$, \ionr{Si}{ii}, \ionr{N}{ii}, \ionr{O}{iii}, \ionr{Si}{iv}, \ionr{O}{vi}, and Fe K$\alpha$. Multiple observations of X-ray emission from nearby star-forming galaxies have been used to estimate key parameters of galactic outflows, such as mass loading factors and thermalization efficiencies etc. \citep{Strickland2000, Strickland2004, HodgesKluck2020, Lopez2020}. All these estimates are, however, heavily dependent on theoretical models for the dynamical and ionization properties of the galactic winds. 

\subsection{Galactic Wind thermodynamics}
Dynamical properties of galactic winds (hereafter, winds) are often described as a simple spherical expansion of the mass and energy injection by the supernovae \citep{Mathews1971, Chevalier1985, MSharma2013b, Thompson2016, Krumholz2017b, Samui2018}. In this description, supernovae deposit mass and energy within a small region in the ISM thus creating a hot and over-pressurized region. This causes the injected material to push the ISM and the CGM significantly away from the star-forming region, thereby creating a steady-state galactic wind within this evacuated region. 

Assuming that the injected energy is in the form of thermal energy, it can be easily shown that the total amount of injected thermal energy quickly becomes kinetic energy not very far from the star-forming region. The maximum velocity of the wind is thus given as $v_w \approx \sqrt{2\:L_{\rm mech}/\dot{M}} \approx 10^3 \sqrt{\epsilon/\beta}\quad \kmps$ (assuming no-gravity), where, $\epsilon$ and $\beta$ are the thermalization efficiency and the mass loading factor of the wind. 
Here, we have defined the total mechanical power and mass outflow rate to be (for consistency with the literature)
\begin{eqnarray}
 L_{\rm mech} &=& \epsilon \,3\times 10^{41} (\mbox{SFR}/\mpy)\, \ergps \nonumber \\
 \dot{M} &=& \beta\, \mbox{SFR}  \,,
 \label{eq:beta-classic}
\end{eqnarray}
corresponding to a star-formation rate, SFR. 
The equations for mass continuity and constant entropy therefore imply that the density, $\rho \propto r^{-2}$ and the temperature, $T \propto r^{-4/3}$ (assuming an adiabatic index $\gamma =5/3$). The temperature cools from $\approx 1.5 \times 10^7 \epsilon/\beta$ K \citep{Chevalier1985} inside the star-forming region to as low as $\sim 10^3$ K as the gas expands. Since radiative cooling efficiency increases with decreasing temperature in this temperature range, the wind dynamics can be further modified by it \citep{Thompson2016}.

Simulations of winds have shown much more complicated dynamics involving acceleration and mixing of cold clouds, creation of entropy due to bow shocks, and non-spherical wind geometries \citep{Cooper2008, Melioli2013, Vijayan2018, Schneider2020}. Theoretical models have been able to include some of these features to some extent \citep{Krumholz2017b, Nguyen2021, Fielding2022}.

\subsection{Wind ionization states}
While better dynamical models for the winds are important, modeling the observations also requires further assumptions about the ionization states of the plasma. The simplest way to connect the dynamical quantities to the ionization states is to assume \textit{ionization equilibrium} at a given temperature and radiation field due to collisional plus photo-ionization. The assumption, however, breaks down if the recombination timescale ($\tau_{\rm rec}$) of the ions is longer than the timescale ($\tau_{\rm th}$) over which the plasma changes its temperature or radiation field. The plasma then enters into a \textit{non-equilibrium ionization} (NEI) phase where the ionization structure of the plasma is not represented by the instantaneous temperature/radiation but is dependent on the history of the evolution \citep{Kafatos1973, Gnat2007, Gnat2017, Sarkar2021b}.

The thermal timescale for winds (assuming purely adiabatic expansion) is 
\begin{equation}
    \tau_{\rm th} \sim \frac{r}{v_w} \approx 0.2\, r_{0.2}\,\sqrt{\frac{\beta}{\epsilon}}\quad \mbox{  Myr}
\end{equation}
where, $r_{0.2} = r/0.2$ kpc. The recombination timescale, on the other hand, is 
\begin{equation}
    \tau_{\rm rec} \sim \frac{1}{n_e\, \alpha_{\rm rec}} 
\end{equation}
where, $n_e$ is the election number density and $\alpha_{ \rm rec}$ is the recombination rate coefficient of an ion \footnote{Here, the recombination rate is the effective recombination rate including radiative + dielectronic recombination rates and collisional ionization rates from the lower state. The effective recombination rate is, therefore, given as $\alpha_{\rm rec} = x_{\rm o ix} \alpha_{\rm o ix}-x_{\rm o viii} \xi_{\rm o viii}$ where, $x_{\rm o viii}$ and $x_{\rm o ix}$ are the ion fractions for \ionr{O}{viii} and \ionr{O}{ix}, and $\alpha_{\rm o ix}$ and $\xi_{\rm o viii}$ are the dielectronic+radiative recombination rate of \ionr{O}{ix} and collisional ionization rate of \ionr{O}{viii}.}
Assuming typical values for a star forming region with size of $200$ pc, star formation rate (SFR) of $10 \mpy$ and say for \ionr{O}{ix} (corresponding to central temperature of $\sim 10^7$ K), the numbers are , $n_e \sim 1\, \pcc$ and $\alpha_{\rm rec} \sim 10^{-14} \cc  \psec$.
The ratio of the recombination time and the dynamical time right before the plasma leaves the star forming region is $\tau_{\rm rec}/\tau_{\rm th} \sim 15$ (assuming, $\epsilon \sim \beta$). The plasma will remain over-ionized compared to the equilibrium values as it expands. 

The ionization states are also affected by the radiation field, especially at low densities and low temperature where collisional ionization is less competitive. The critical density below which photoionization dominates is
\begin{equation}
    n_{\rm e, crit} = \frac{\Gamma_i (J) }{\xi_i}
\end{equation}
where, $\Gamma_i$ is the photoionization rate at any given radiation flux $J(\nu)$ and $\xi_i$ is the collisional ionization rate coefficient for any ion $i$. As an example, $\Gamma_i = 2.2 \times 10^{-15} \psec$ and $\xi_i = 2\times 10^{-13} \psec \cc$ for \ionr{O}{iii} at a temperature of $6\times 10^6$ K and in the presence of the metagalactic radiation background \cite[][hereafter, HM12]{Haardt2012}. This implies that \ionr{O}{iii} is going to be photo-ionized in the mere presence of the HM12 radiation for gas densities $\lesssim 10^{-2} \pcc$ easily realized in the expanding winds. 
These arguments suggest that we need to consider proper ionization dynamics considering NEI and photo-ionization (PI) in addition to the thermodynamics of the winds.

\subsection{The current paper}
Despite being of significant importance there have been only a few efforts to quantify the NEI physics and radiation field in galactic winds. \cite{Breitschwerdt1994} estimated that the NEI physics in winds from superbubbles near the Sun can produce significant X-ray emission that dominates the X-ray background at Earth. In recent work, \cite{Gray2019} performed simulations of galactic winds including the effect of NEI and radiation from the newly formed O/B stars and metagalactic radiation background. They found that the NEI process can alter the ionization states of the plasma by many orders of magnitude.
\cite{Oskinova2022} showed the presence of an excess \ionr{He}{ii} ionizing radiation from the hot gas in superbubbles but did not include the radiation from the star cluster.

In this paper, we perform simulations of spherically expanding winds from a star-forming region (similar to \cite{Gray2019}) but include frequency-dependent radiative transfer to consider the self-radiation from the wind itself, in addition to the UV radiation from the young stars and the HM12 UV/X-ray background. We show that the X-rays produced by the hot wind itself can keep the wind self-ionized. This excess radiation can be a factor of few to a few orders of magnitude higher compared to the HM12 background and, therefore, is extremely crucial for modeling warm/cold gas in/around galactic winds. In this paper, we do not include any multiphase hydrodynamical interactions between the fast wind and cold clouds or departure from spherical symmetry as in \cite{Fielding2022, Nguyen2021}.

\section{Numerical techniques}
\label{sec:numerical-technique}
Most of the numerical packages used in this paper have been described in detail in \cite{Mignone2007, Tesileanu2008, Sarkar2021a} \footnote{The modifications to the \pluto-4.0 code are publicly available at \url{https://gitlab.com/kartickchsarkar/pluto-neq-radiation}}. However, we provide a brief description as well as new improvements and simulations set up for the current context in the following sections. 

\subsection{Hydrodynamic code}
\label{subsec:hydro-code}
We use the Eulerian grid code \textsc{pluto}-v4.0 \citep{Mignone2007} that uses Godunov scheme to solve the Riemann problem in fluid dynamics. We assume spherical symmetry for the fluid and ionization components. The fluid equations under consideration are 
\begin{equation}
\label{eq:mass-eq}
\ddt \rho + \frac{1}{r^2} \ddr(r^2 \rho v) = \dot{\rho}_{\rm sn}
\end{equation}
\begin{equation}
\label{eq:mom-eq}
\ddt (\rho v) + \frac{1}{r^2} \ddr(r^2 \rho v^2) = -\ddr p - \ddr \phi + \rho a_r
\end{equation}
\begin{equation}
\label{eq:energy-eq}
\ddt (E+\rho \phi) + \frac{1}{r^2} \ddr \left[r^2\:(E+p+\rho \phi) v\right] =\dot{E}_{\rm sn} + \mathcal{H}-\mathcal{L} + \rho v a_r 
\end{equation}
where, $\rho, v$ and $p$ are the mass density, the radial velocity and the thermal pressure, $E = (1/2) \rho v^2 + p/(\gamma-1)$ is the total energy density, $\phi$ is the gravitational potential, $\dot{\rho}_{\rm sn}$ is the mass injection rate from core-collapse SN, $a_r$ is the radial acceleration due to the local radiation field, $\dot{E}_{\rm sn}$ is the thermal energy injection rate within the star forming region, $\mathcal{H}$ is the heating rate, and $\mathcal{L}$ is the radiative cooling rate. The heating and cooling rates are calculated consistently from the local radiation field and non-equilibrium ion-fractions of the plasma. The calculation of the source terms are described in the following sections.

\subsection{NEI code}
The hydrodynamics of the ions are solved using the advection equation assuming that the ions do not contribute to the dynamics of the fluid directly and simply flow along with the fluid \citep{Tesileanu2008}. We consider all ionization states of H, He, C, N, O, Ne, Mg, Si, S, and Fe in our calculation. The abundances for the elements are assumed to be Solar and taken from \cite{Asplund2009}. 

The equations for the ion dynamics, in the conservative form, are
\begin{equation}
    \label{eq:nei-eq}
    \ddt (\rho x_{k,i}) + \frac{1}{r^2} \ddr(r^2 \rho x_{k,i} v) = \rho S_{k,i}
\end{equation}
where, $S_{k,i}$ is the rate of change of ion fraction, $x_{k,i}$, of an ion $i$ belonging to an element $k$, due to ionization and recombination, and is given as 
\begin{eqnarray}
    S_{k,i} &=& n_e \left[x_{k,i-1} \xi_{k,i-1} - \left(\alpha_{k,i}+\xi_{k,i}\right) x_{k,i} + \alpha_{k,i+1} x_{k,i+1}\right] \nonumber \\
    &-& x_{k,i} \Gamma_{k,i} + A_{k,i}
    \label{eq:S_ki}
\end{eqnarray}
Where, $\xi_{k,i}$ is the collisional+charge transfer ionization rate coefficient from ion $(k,i)$ to $(k,i+1)$, $\alpha_{k,i}$ is the radiative + dielectronic+charge transfer recombination rate from ion $(k,i)$ to $(k,i-1)$, $\Gamma_{k,i}$ is the PI rate of an ion $(k,i)$ in the presence of the local radiation field $J_\nu$, and $A_{k,i}$ is the Auger ionization rate from $(k,<i)$ to $(k,i)$ and is calculated by summing up all the contributions of lower states to eject 1 or more electrons.

Cooling rates are calculated based on the local non-equilibrium ion fractions of elements and include radiative cooling as well as the cooling due to charge transfer. For radiative cooling, we use ion-by-ion cooling rates  \cite[following][]{Gnat2012} and the charge transfer rates from \textsc{cloudy-17} \citep{Cloudy2017}. The heating rate is also consistently calculated from the instantaneous ion fractions and the radiation field, and also includes heating due to charge transfer (see \cite{Sarkar2021a} for more details).

\begin{table*}
    \centering
    \caption{Values and their meanings for the parameters in the galaxy gravitational potential, following \protect\cite{Sarkar2017}.  $^\dag$ Note that in the current format in eq  \ref{eq:grav-potential}, the potential, and therefore, the density actually represents the distribution along the minor axis of the galaxy.}
    \begin{tabular}{c|c|c}
    \hline\hline
    Mass parameter & description & value \\
    \hline
        $M_{\rm bulge}$ & Mass of the bulge & $2\times 10^{10} \msun$ \\
        $M_{\rm disk}$ & Mass of the disk & $6\times 10^{10} \msun$ \\
        $M_{\rm vir}$ & virial mass of the galaxy & $1.2\times 10^{12} \msun$ \\ 
        $r_b$ & bulge scale radius & $2$ kpc\\
        $a,b$ $^{\dag}$ & stellar disk scale radius and height & $3, 0.4$ kpc \\
        $r_c$  & dark matter core radius & 6 kpc\\
        $r_s$ & dark matter scale radius & 21 kpc\\
        $c$ & dark matter concentration parameter & 12 \\
        $m(c)$ & - & $\log (1+c) - c/(1+c)$ \\
        \hline\hline
    \end{tabular}
    \label{tab:mass-params}
    \vspace{0.5cm}
\end{table*}

\subsection{Radiative transfer}
We consider the time-independent radiative transfer equation in spherical symmetry,
\begin{equation}
    \label{eq:rad-transf}
    \frac{\mu}{r^2} \ddr (r^2 \psi_\nu) + \frac{1}{r}\: \frac{\partial}{\partial \mu} \left( (1-\mu^2) \psi_\nu \right) = j_\nu - \alpha_\nu \psi_\nu 
\end{equation}
where, $\psi_\nu \equiv \psi(r,\mu, \nu)$ is the specific intensity, $\mu = \cos\theta$ is the cosine of the angle between a ray and the radial direction, $\nu$ is the frequency, $j_\nu$ is the local volume emissivity and $\alpha_\nu$ is the local absorption coefficient. This equation is solved for $56$ frequency bands (starting from 5 eV to 2.5 keV) that considers major ionization edges of H, He and metals. The calculation of the volume emissivity and absorption coefficient also takes the local non-equilibrium ionization states into account. While ion-by-ion cooling rates provide the bolometric emission from any interaction, ion-by-ion emissivity requires further distinction in frequency as well as with electron density. We have compiled a library of spectra for each ion with varying temperature and electron density using \textsc{cloudy}-17.

We use the method of short characteristics to solve for the radiative transfer equation. This method solves $\psi_\nu$ on a fixed grid of $\mu$ at each radial grid. To conserve the flow of radiation energy across the $r-\mu$ grid while discretizing equation \ref{eq:rad-transf}, we use a `finite volume' method (discretization by integration) instead of the more often used `finite element' method (discretization by differentiation). 

Finally, for a complete solution, we solve eq \ref{eq:mass-eq}, \ref{eq:mom-eq},  \ref{eq:energy-eq}, and \ref{eq:nei-eq} in a coupled manner. Eq. \ref{eq:rad-transf} is solved at the last stage of the integration for the new $\rho, p$ and ion fractions. The solved radiation field is then used for the next hydrodynamic step. These techniques and validations are described in detail in \cite{Sarkar2021a}.

\begin{figure*}
    \centering
    \includegraphics[width=\textwidth]{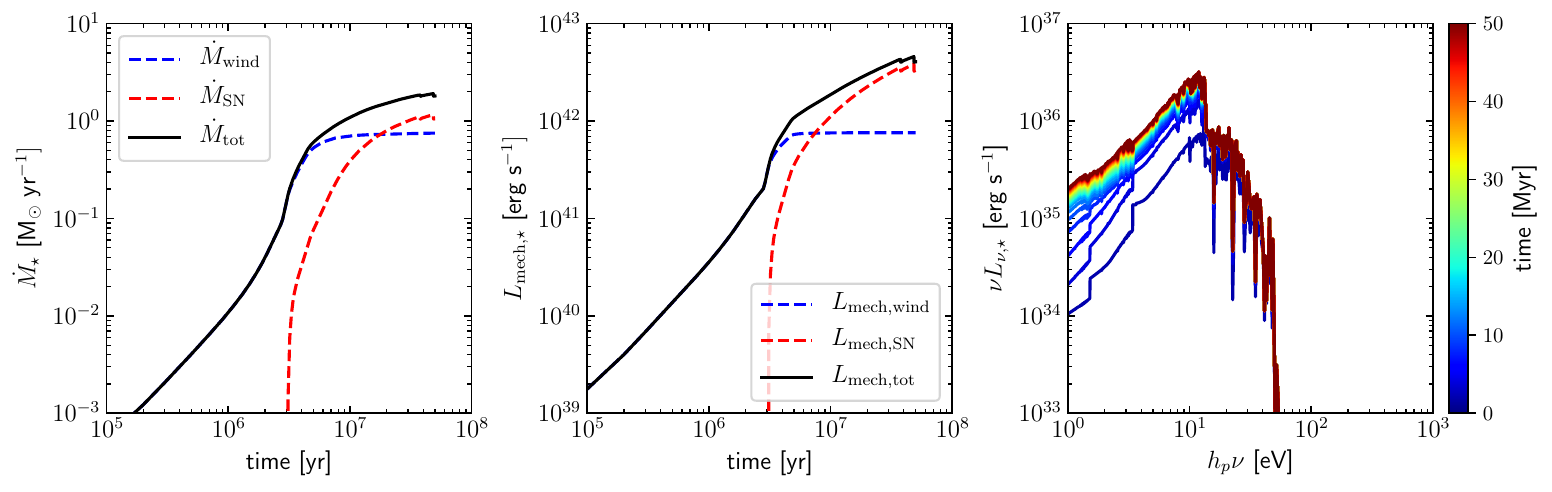}
    \caption{Total mass (left panel), mechanical power (middle panel) and radiation energy (right panel) output from the stellar synthesis code \textsc{starburst99} for a given star-formation rate of $10 \mpy$. The input radiation spectra in the simulation uses a coarser frequency containing only 56 frequency bins than shown here. The injection rates become roughly constant after $\sim 30$ Myr when the wind attains a steady-state solution.}
    \label{fig:sb99_input}
    \vspace{0.5cm}
\end{figure*}

\subsection{Initial conditions}
\label{subsec:boundary-condition}
Our simulation box extends from $20$ pc to $10$ kpc with a spatial resolution of $1.2$ pc. The initial density distribution is assumed to be in hydrostatic equilibrium with the background gravitational potential, $\Phi(r)$, provided by the stellar bulge and disk, and the dark matter. The density distribution of the ambient as is given as 
\begin{equation}
    \rho_a(r) = \rho_a (0) \exp\left[ - \frac{\bar{\mu} m_p}{k_B T_a} (\Phi(r)-\Phi(0)) \right]
\end{equation}
where, $T_a = 2\times 10^6$ K is the assumed ambient temperature, $\bar{\mu} = 0.6$ is the mean molecular weight, and $\rho_a(0) = 10^{-3} \mpcc$ is the central density of the hot ambient CGM. The background potential is given as \citep{Miyamoto1975, NFW1997}
\begin{eqnarray}
\Phi(r) &=& \Phi_{\rm bulge}(r) + \Phi_{\rm disk}(r) + \Phi_{\rm nfw}(r) \nonumber \\
&=& -\frac{G M_{\rm bulge}}{\sqrt{r^2+r_b^2}} - \frac{G M_{\rm disk}}{a+\sqrt{r^2+b^2}} \nonumber \\
&& - \frac{G M_{\rm vir}}{r_s\:m(c)} \frac{\log\left( 1+ \frac{\tilde{r}}{r_s}\right)}{\tilde{r}/r_s}  \,.
\label{eq:grav-potential}
\end{eqnarray}
where, $\tilde{r} = \sqrt{r^2+r_c^2}$ is a modified radius to avoid a cuspy dark matter density profile. The values and descriptions of the parameters are given in Table \ref{tab:mass-params}.

The radiation field at the outer boundary ($r= 10$ kpc) is set to be equal to the HM12 background at all $\mu < 0$. The inner boundary ($r=20$ pc) is set to be reflective (by construction, for the radiative transfer solver). The initial ionization states are assumed to be everywhere in collisional+photo equilibrium given the temperature of the ambient medium and the HM12 background. 

\subsection{Mass, energy and radiation injection}
\label{subsec:energy-inj}
We assume that the star formation is uniformly distributed within a spherical radius, $R = 200$ pc (also referred to as the injection region). Although this value is chosen arbitrarily, it is consistent with the observed size of the star-forming region in the nearby galaxy M82 \citep{Strickland2000}. We  assume a constant star-formation rate (SFR) starting at $t=0$  and throughout the simulations. The mass and energy injection is initially dominated by the stellar wind feedback from the young and massive O/B stars. It is only after $\sim 3$ Myr that the mechanical energy from supernovae becomes important. The bolometric luminosity from the star-forming region also evolves as the stellar population approaches equilibrium  \citep[see e.g.][]{Sternberg2003a}. To consider a complete time evolution of the energy, we assume the mass, energy, and radiation output from the stellar synthesis code \textsc{starburst99} \citep{Leitherer1999, Vazquez2005, Leitherer2010, Leitherer2014} as a function of time. 

\textsc{starburst99} is run for a constant SFR of $1 \mpy$. The values at any given time are obtained by interpolating the output. For the stellar population, we assume i) Solar metallicity, ii) Kroupa initial mass function with broken power-law index of $1.3$  from $0.1 \msun$ to $0.5 \msun$ and  $2.3$ from $0.5 \msun$ to $100 \msun$, and iii) supernova cutoff mass $=8 \msun$. The outputs for the other SFR cases are obtained from simply multiplying by the SFR to these outputs. While the mass and energy outputs are obtained at an interval of $0.1$ Myr, the spectra output is obtained at a time interval of $2$ Myr. The mass ($\dot{M}$) and mechanical energy ($L_{\rm mech}$) injection from the stellar wind and SNe are converted into a density and internal energy addition rate following 
\begin{eqnarray}
    \dot{\rho}_{\rm sn} &=& \frac{3\dot{M}}{4\pi R^3} = \eta_M  \frac{3\dot{M}_\star}{4\pi R^3} \nonumber \\
    \dot{E}_{\rm sn} &=& \frac{3 L_{\rm mech}}{4\pi R^3} =  \eta_E  \frac{3 L_{{\rm mech},\star}}{4\pi R^3} \,.
\end{eqnarray}
where, $\eta_M$ and $\eta_E$ are the fractions of mass and energy that the injected material gains or loses due to either mass loading or radiative losses. Considering the typical values from \textsc{starburst99} at $t\sim 30$ Myr (see fig \ref{fig:sb99_input}), the traditional mass loading factor and thermalization efficiency (see eq \ref{eq:beta-classic}) are  
\begin{eqnarray}
\beta &\approx& 0.2\, \eta_M\nonumber \\
\epsilon &\approx& 1.3\, \eta_E 
\label{eq:eta-beta}
\end{eqnarray}

The addition of the stellar radiation energy density is a bit complex since the radiation field is not a scalar and originates from point sources. However, one can define an equivalent emissivity assuming that these radiation sources are placed randomly within the injection region such that the radiation flux at $r=R$ is recovered. This equivalent emissivity is given as (see appendix \ref{app-sec:equiv-emissivity})
\begin{equation}
    \varepsilon_{\nu,\star} = \frac{3 L_{\nu,\star}}{4\pi R^3}\,,
    \label{eq:equiv-emissivity}
\end{equation}
where, $L_{\nu,\star}$ is the luminosity output from \textsc{starburst99}, for a given star formation rate. Unlike the mass and energy injection, we do not assume any loss of radiation field due to absorption since the central region is mostly very hot ($\sim 10^7$ K) and therefore optically thin to most of the stellar radiation. The input radiation field from the massive stars is shown in the right panel of figure \ref{fig:sb99_input}. It shows that the stellar radiation increases over time and finally reaches a maximum at $t\gtrsim 20$ Myr. The radiation field is, however, only dominant below $\lesssim 60$ eV, characteristic of the temperature of the most massive O stars \citep{Sternberg2003a}. 

\subsection{Parameters and runs}
\label{subsec:table-of-runs}
The current paper is exploratory in nature to study the effects of the new physics in the winds, especially self-radiation. We, therefore, restrict ourselves to studying only one case for a single set of parameters as listed in Table \ref{tab:wind-params}. We choose the parameters based on known values of the nearby star-forming galaxy M82 \citep{Strickland2000, Strickland2009}.

To study the effect of the various physical effects in the galactic wind, we run four different simulations.
First, \pie (photo+collisional ionization equilibrium) is the simplest and does not include the non-equilibrium ionization network. The ionization fractions and cooling are calculated assuming collisional+photo equilibrium for the given density, temperature, and the radiation field from the HM12 background and stellar radiation field (following a proper radiative transfer). 
Second, \nei includes non-equilibrium ionization and calculates the ion fractions by solving equation \ref{eq:nei-eq} in the presence of a local radiation field. Third, \piesr (photo+collisional ionization equilibrium and self-radiation), is the same as \pie but now includes the radiation produced by the plasma itself. Fourth, \neisr combines the NEI effects and self-radiation. Our four simulation runs are summarized in Table \ref{tab:runs}.
\begin{table}
    \centering
    \caption{Parameters for the star-forming region and the wind. The values roughly follow the estimated values for the M82 \protect\citep{Strickland2009}.}
    \begin{tabular}{c|c|c}
    \hline \hline
    parameters & description & value \\
    \hline 
    SFR & star-formation-rate & $10 \mpy$ \\
    R   & star-forming region radius & $200$ pc \\
    $\eta_M$ & mass loading efficiency & $3$ \\
    $\eta_E$ & energy loading efficiency & $0.3$ \\
    \hline\hline
    \end{tabular}
    \label{tab:wind-params}
\end{table}

\begin{table}
    \centering
    \caption{Different simulations run in this paper. The simulations vary in terms of their ionization physics and radiation field. The equilibrium ionization is calculated considering the collisional ionization, recombination, and photo-ionization due to the local radiation field. The radiation field is itself calculated by solving a full radiative transfer equation as discussed in section \ref{sec:numerical-technique}.}
    \begin{tabular}{c|c}
    \hline \hline
    Simulations & Physics included \\
    \hline
    \pie & hydrodynamics, \\
         & radiation from stars+HM12, \\
         & photo + collisional ionization equilibrium\\
         \hline
    \nei & hydrodynamics, \\
         & radiation from stars+HM12, \\
         & non-equilibrium ionization \\
         \hline
    \piesr & hydrodynamics,\\
           & radiation from stars+HM12+wind\\
           & photo+collisional ionization equilibrium \\
           \hline
    \neisr & hydrodynamics, \\
        & radiation from stars+HM12+wind,\\
        & non-equilibrium ionization \\
    \hline\hline
    \end{tabular}
    \label{tab:runs}
\end{table}
%
%
%
%
\section{results}
\label{sec:results}

\begin{figure*}
    \centering
    \includegraphics[width=\textwidth]{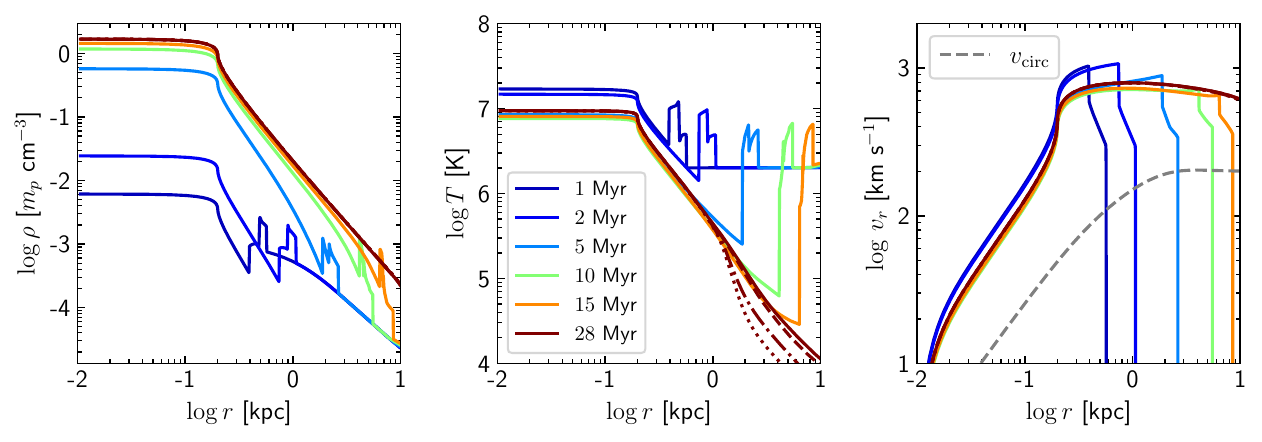}
    \caption{Evolution of thermodynamic states of the wind for the \neisr run. The density, temperature, and velocity of the wind are shown in the left, middle, and right panels. The mass and energy injection push the background CGM further and further with time and finally set up a steady-state flow at $\gtrsim 20$ Myr. The injection region is at $R\leq 200$ pc where the density and temperature are roughly constant. Different line-styles show results for different runs in the steady state (dotted:\pie, dash-dotted: \piesr, dashed:\nei, and solid: \neisr). The lines superpose each other except for the temperature panel. The dashed gray line in the right panel shows the circular velocity corresponding to the assumed gravitational potential.}
    \label{fig:wind-thermodynamics}
    \vspace{0.5cm}
\end{figure*}
%
\subsection{Thermodynamic evolution}
\label{subsec:wind-thermodynamics}
The thermodynamic evolution of the wind is shown in Fig \ref{fig:wind-thermodynamics}. We inject mass, energy, and radiation output from SN and massive stars into the star-forming region, $R$ (also called as the injection region). Once the injection region accumulates enough  energy, it drives a strong shock through the ambient CGM. With time, the shock sweeps away the ambient CGM and finally sets up a steady-state flow within the simulation box at $t\gtrsim 20$ Myr. Throughout this time, both the density and temperature of the injection region rise due to the increase in injection rate from the growing numbers of stars and SNe (see figure \ref{fig:sb99_input}) plateauing at $\gtrsim 25$ Myr. The steady-state flow roughly maintains the same density, velocity and temperature profiles after this time.  

The wind follows a pure adiabatic expansion in the absence of radiative cooling. 
Enhanced cooling via radiative losses sets in at the cooling radius, $r_{\rm cool}$, when the cooling time, $\tcool$ becomes smaller than the expansion time $\tdyn \sim r/v_r$ \citep{Mathews1971, Thompson2016}. We find this radius to be about $2$ kpc in the steady state
(middle panel of figure \ref{fig:wind-thermodynamics}; for $\eta_E = 0.3\,,\eta_M = 3$ and $\Omega = 4\pi$), consistent with the estimate given by equation 6 of \citep{Thompson2016}. The presence of non-equilibrium physics and radiation, however, changes the temperature profile at $r\gtrsim r_{\rm cool}$. Figure \ref{fig:wind-thermodynamics}, also compares the thermodynamic profiles of the wind for different runs in the steady state. The wind is susceptible to over-ionization which suppresses the radiative cooling efficiency as shown by the dashed line in the temperature panel. The presence of excess radiation from the wind (see sec \ref{subsec:Radiation-field}) also heats the wind such that radiative cooling is delayed (shown by dash-dotted line). The addition of both the NEI physics and self-radiation can suppress cooling to an extent where the temperature profile is very close to the adiabatic profile. However, we stress that the amount of cooling suppression is dependent on the wind parameters, $\epsilon,\,\beta$, SFR, and  $R$.  

The density and velocity profiles, on the other hand, are not affected by this additional cooling (or other physics included in this paper) since the wind has already converted most of its thermal energy into kinetic energy by this radius (since $r_{\rm cool} \gg R$). The wind, however, slows down by a bit at $r \sim 10$ kpc due to the gravitational forces from the galaxy. The decelerated wind velocity at $r\gg R$ can be estimated, using the momentum conservation equation, to be
\begin{equation}
    v(r)^2 \approx v_w^2 - 2 v_{\rm circ}^2(r)\, \log\left(\frac{r}{R}\right)\,,
    \label{eq:vr_vw}
\end{equation}
where, $v_{\rm circ}(r) = \sqrt{r\: d\Phi(r)/dr}$ is the circular velocity at $r\gg R$ \citep[see equation 6 of][]{Fielding2022}. The circular velocity in our simulations 
is shown by the dashed gray line in the right panel of figure \ref{fig:wind-thermodynamics}. 
The above equation shows that the wind velocity 
decreases from $v_w \approx 800 \kmps$ at $300$ pc to $\approx 600 \kmps$ at 10 kpc for an assumed $v_{\rm circ} \approx 200 \kmps$, consistent with the simulations. 

In the current paper, we will mostly focus on the steady state behavior of the wind, i.e. at $\gtrsim 20$ Myr for our simulations. 

%
\begin{figure*}
    \centering
    \includegraphics[width=0.8\textwidth]{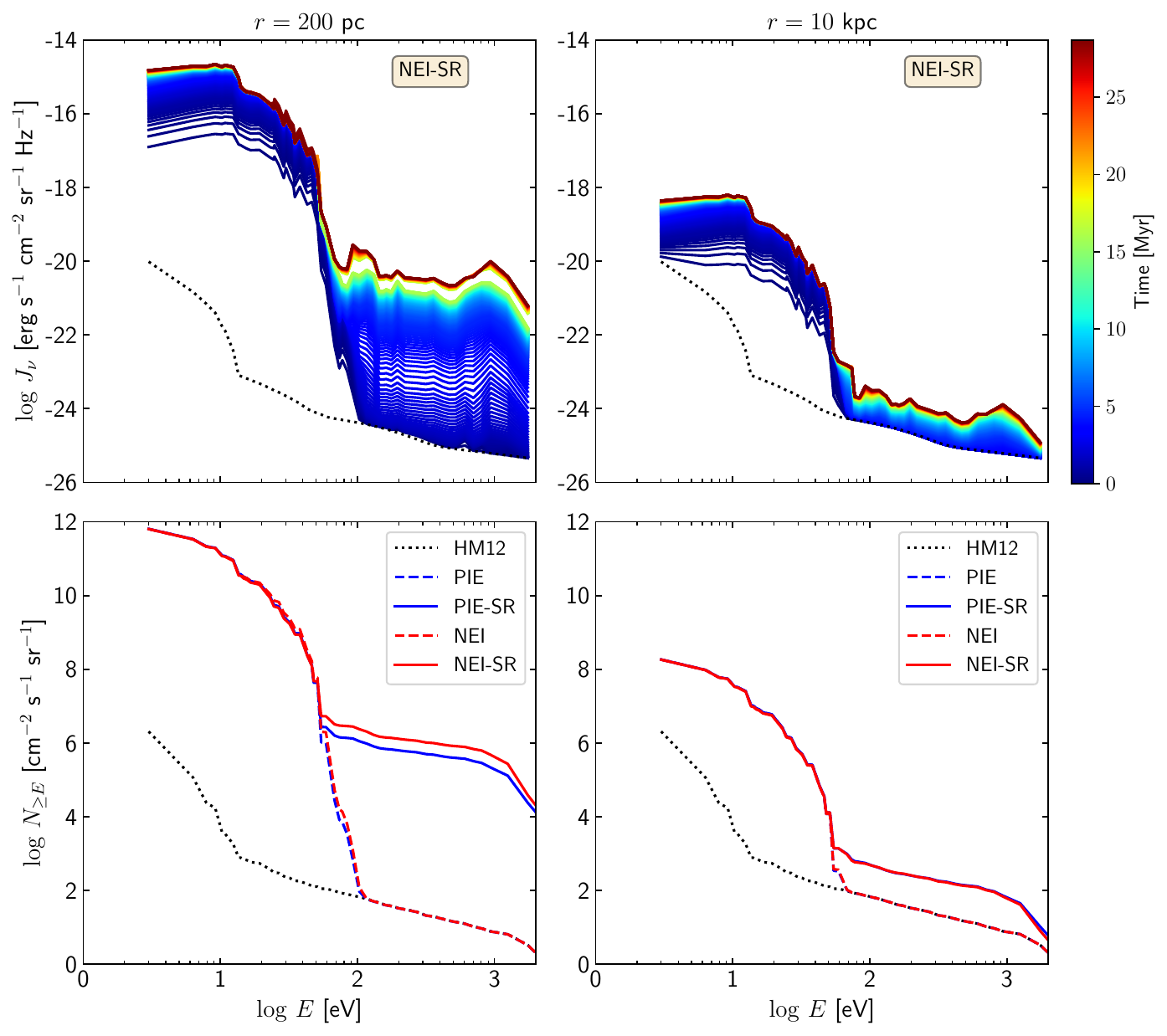}
    \caption{Output spectra at $200$ pc (left panel) and at $10$ kpc (right panel). The \textit{top panel} shows the spectral evolution with time for the \neisr run. Different lines. The \textit{bottom panel} compares the photon number-spectra (total number of photons above certain energy) from all the four runs at $t = 28$ Myr. The black dotted lines represent the HM12 background radiation field. The dashed lines represent runs without considering self-radiation of the wind itself and the solid lines show results for the cases with self-radiation of the wind included.}
    \label{fig:spectra-nphots}
    \vspace{0.5cm}
\end{figure*}

\subsection{Radiation field}
\label{subsec:Radiation-field}
As figure \ref{fig:wind-thermodynamics} shows, the gas particle density can vary from $\sim 1$ to $10^{-4} \pcc$ inside the wind and the temperature from $\sim 10^7$ to $\sim 10^4$ K thus producing ions that can be observed from X-ray to optical/UV wavebands. While the density and temperature can qualitatively indicate the kind of ions at any region, the proper populations depend on the thermal history of the plasma (non-equilibrium ionization) and the local radiation field. In this subsection, we study the radiation field in the wind. 

Figure \ref{fig:spectra-nphots} shows the radiation field at two locations, one, just outside the star-forming region ($r=R=200$ pc) and the second, at the edge of the simulation box ($r=10$ kpc) for the \neisr run. The top panel shows the evolution of the radiation spectra (angle averaged specific intensity; $J_\nu$) at these two locations. We find that $J_\nu$ is dominated by the radiation from massive stars only at $E \lesssim 100$ eV (soft radiation), as expected (see discussion in section \ref{subsec:energy-inj}). At higher energies ($E \gtrsim 100$ eV; hard radiation), the field is dominated by the HM12 background radiation field only at early times when the wind has not yet affected the CGM in the simulation box. At later times, we find that the hard radiation field increases with time and reaches a steady-state value which is almost $4-5$ orders of magnitude higher (at $r = R$) compared to the HM12 field. This excess of hard radiation decreases with radius but still remains $\sim 1$ order of magnitude higher at $r = 10$ kpc. The origin of this excess hard-radiation field is the emission from the wind itself, as will be clear below. 

The bottom panel of figure \ref{fig:spectra-nphots} shows the photon number-spectra ($N_{\geq E}$; the total number of photons above energy $E$) for all the runs at $t = 28$ Myr when the wind has reached a roughly steady-state thermodynamic profile. The dashed lines represent the runs with self-radiation turned off, whereas, the solid lines represent the runs with the self-radiation turned on. The dotted line shows the HM12 background radiation (at redshift zero) for comparison. In the absence of self-radiation, the soft radiation is dominated by the radiation from the stars and does not depend on the non-equilibrium physics, whereas, the hard-radiation is dominated by the HM12 background, as expected. The hard radiation field, however, is much higher in the cases with self-radiation turned on. 
It is, therefore, evident that the excess hard radiation is due to the emission from the wind itself. As far as the spatial origin of this radiation is concerned, we see in figure \ref{fig:wind-thermodynamics} that the wind is sufficiently hot ($\sim 10^7$ K) and dense ($\sim 1\,\pcc$) inside the injection region to produce hard-radiation at $E\gtrsim 100$ eV. The wind also remains hot and dense outside this region but cools down quickly with radius and thus produces only a small amount of radiation compared to the plasma in the injection region. We also notice that the presence of NEI physics makes the hard-radiation field just a factor of a few higher compared to a pure ionization equilibrium case. This is due to the fact that NEI allows higher ions (such as \ionr{O}{viii} and \ionr{O}{ix}) to remain over-ionized compared to their ionization equilibrium values (at that temperature) and thus produce more hard-radiation via free-bound emission processes.

\subsection{Modeling excess radiation field}
\label{subsec:excess-rad-model}
Now, since the density and the temperature of the wind depends on the SFR, $\eta_M$, $\eta_E$, and $R$, the excess hard-radiation is also a function of these wind parameters.
A simple estimate of the total bolometric luminosity for the star-forming region assuming density and temperature to be constant, is expected to follow \citep[see eq 3 of][]{Sarkar2016}
\begin{equation}
    L_{X,C} \approx 3\times 10^{39} \epsilon^{-1} \beta^3 \left(\frac{\mbox{SFR}}{\mpy}\right)^2\, R_{100 {\rm pc}}^{-1} \Lambda_{-23}\quad \ergps
\end{equation}
where $\Lambda_{-23} = \Lambda(T)/10^{-23} \ergps \cc$ is the cooling function, $R_{100 {\rm pc}} = R/100$ pc. The specific intensity at $r=R$ is then given by $\psi_X (\mu) = 3 \mu L_{X,C}/(8\pi^2 R^2)$  for $\mu \geq 1$ and $=0$ otherwise (see eq \ref{app-eq:stellar-psi}). The angle averaged specific intensity (comparable to the bottom panel of fig \ref{fig:spectra-nphots})  at any given radius, $r$, is 
\begin{eqnarray}
N_X (r) &=& \frac{1}{2} \int_{\mu_c}^{+1} \psi_X (\mu) d\mu \times  \frac{1}{\bar{E}}\nonumber \\
   &=& \frac{3\, L_{X,C}}{32\:\pi^2 r^2 \times \bar{E}} \nonumber \\
   &=&  \frac{1.8\times 10^5\, \beta^3}{\epsilon\,R_{100 {\rm pc}}^3\, \bar{E}_{\rm keV}} \left(\frac{SFR}{\mpy}\right)^2\,  \Lambda_{-23} \times \left(\frac{R}{r} \right)^2 \nonumber \\
   && \quad \photps \pcmsq \mbox{ sr}^{-1}
   \label{eq:N_XC}
\end{eqnarray}
where, $\bar{E} = \bar{E}_{\rm keV}$ keV is the average photon energy of the emitted photon and $\mu_c = \sqrt{1-R^2/r^2}$ is the cosine of the maximum angle that the star-forming region subtends at $r$. Now, considering $\bar{E} \sim$ keV (top panel of figure \ref{fig:spectra-nphots}), $\Lambda_{-23} \approx 2$ and the wind parameters in table \ref{tab:wind-params}, the above equation translates to a photon flux at $r=R = 200$ pc to
be $N_X (R) \sim 2 \times 10^6 \photps \pcmsq$ sr$^{-1}$, a factor
of few larger but consistent with the value ($5\times 10^5 \photps \pcmsq$ sr$^{-1}$) shown in the bottom panel of figure
\ref{fig:spectra-nphots}. This overestimation is because we assumed that all the photons have the same energy. Ideally, there will be a significant fraction of the energy emitted at low energies too.

The exact value of the central luminosity, and hence the photon flux, depends on the profile of the wind (inside the injection region as well as outside) and departure from equilibrium. Here we consider wind parameters consistent with the observations of M82 \citep{Strickland2009}.

\begin{figure*}
    \centering
    \includegraphics[width=0.8\textwidth]{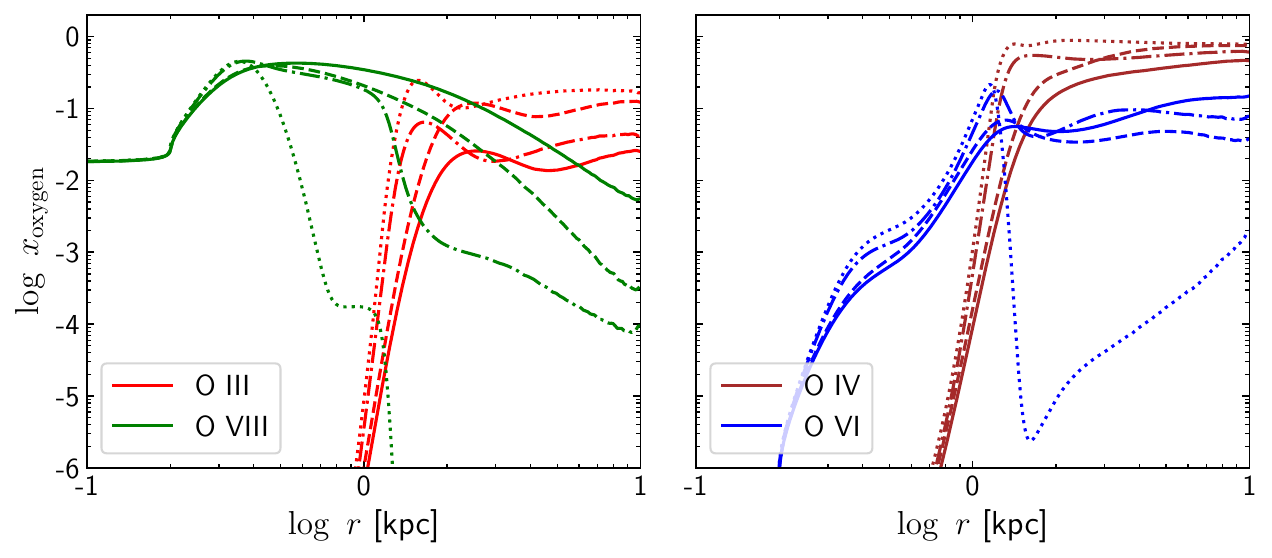}
    \caption{Oxygen ion fractions for different runs at $28$ Myr, when the wind has reached a steady state. The left panel shows ion fractions for \ionr{O}{iii} and \ionr{O}{viii}, whereas, the right panel shows results for \ionr{O}{iv} and \ionr{O}{vi}. Different line-styles show results for \pie (dotted), \piesr (dash-dotted), \nei  (dashed), and \neisr (solid) runs. As a general mnemonic, included physics increases going from dotted to solid curves.}
    \label{fig:O-ionfracs}
\end{figure*}
\begin{figure*}
    \centering
    \includegraphics[width=0.8\textwidth]{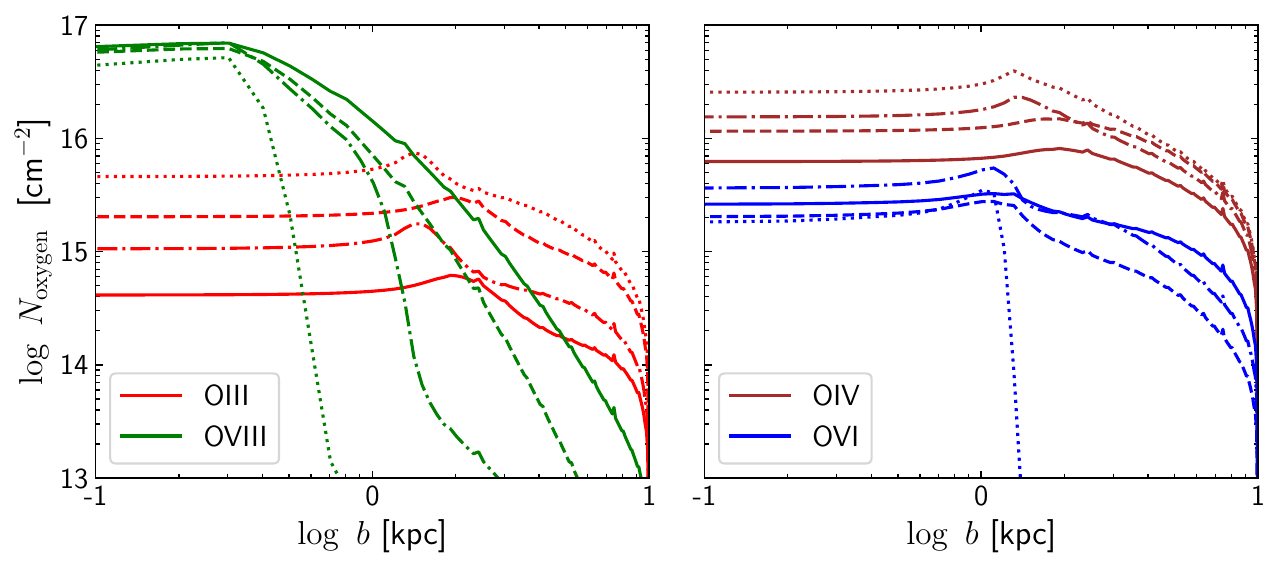}
    \caption{Column density for Oxygen ions at $t=28$ Myr as a function of the impact parameter ($b$) from the center of the star-formation injection region. Different line-styles represent different simulations:
    \pie (dotted), \piesr (dash-dotted), \nei (dashed), and \neisr (solid).}
    \label{fig:colden-oxygen}
    \vspace{0.5cm}
\end{figure*}
\subsection{Ion fractions}
\label{subsec:ion-fracs}
The effects of NEI and excess hard-radiation are best judged by their effects on the elemental ion fractions. After all, most of the observations probe the metal emission and absorption rather than the temperature and density directly. For a qualitative understanding of the effects, we show the ion-fractions of several Oxygen ions in figure \ref{fig:O-ionfracs} for different runs. As a general trend, the rise in ion fractions follows the temperature. For example, \oviii rises to a maximum at $r\sim 400$ pc when $T\sim 2\times 10^6$ K, similarly, \oiii rises to a maximum at $r \sim 2$ kpc when $T \sim 10^5$ K, and so on. 
The evolution of the ions afterward, however, depends on the physics included in the simulations. The evolution of lower ions such as \oiii (ionization potential $\approx 48$ eV) and \oiv (ionization potential $\approx 75$ eV) are heavily dependent on the radiation field as can be seen in \pie run. If the lower ions were only collisionally ionized, then these ion fractions would be peaked since the temperature varies rapidly with radius. However, the addition of stellar plus the HM12 background radiation (run \pie) maintains higher ion fractions for these ions via photo-ionization even after the plasma temperature falls to values where the CIE fractions would be negligible. 
The intermediate ion \ovi (ionization potential $\approx 138$ eV) mostly follows the temperature at $r \lesssim 2$ kpc in the \pie run but slowly rises at larger radii and increases to $\sim 10^{-3}$ at $10$ kpc. This rise in \ovi is due to the decreasing density with radius that raises the ionization parameter (ratio of photon density to the ion density; see figure \ref{app-fig:ioniz-param}). Higher ions, such as \oviii (ionization potential $\approx 871$ eV) are not affected by the presence of the HM12 background since the number of available ionizing photons is small. 

The above discussion suggests the importance of the radiation field in ionizing the wind. We now turn our attention to our \piesr run where we explicitly include the novel effects of the wind-self-radiation. The excess hard self-radiation(see sec \ref{subsec:Radiation-field}) not only alters the ion-fractions for the lower and intermediate ions, it also affects high ions such as \oviii. Figure \ref{fig:O-ionfracs} shows that \oiii is reduced by almost an order of magnitude, conversely, \ovi is increased by a few orders of magnitude higher compared to the \pie case, and is close to photo-equilibrium with the self-radiation. The figure also shows that the excess hard-radiation can also keep \oviii photo-ionized to $\sim 1$ kpc, after which the ion fraction declines, although remains still a few orders of magnitude higher compared to the \pie case.  

NEI physics is important for ions since the effective recombination times are much longer compared to the wind expansion time. It is therefore expected that the higher ions will be over-ionized as the wind expands. This is exactly what the \nei run shows (dashed lines in figure \ref{fig:O-ionfracs}). We see that the \oviii fraction starts from $\sim 10^{-2}$ at $r<R$ and rises to $\approx 0.5$ at $r\approx 400$ pc following the temperature of the plasma. However, \ovii does not fall immediately after this peak, but rather stays high due to the long recombination time. The \ionr{O}{vii} becomes negligible at $r\gtrsim 10$ kpc. Intermediate ions, such as \ovi also follows the same trend of delayed recombination and maintains a significantly high ion fraction at $10$ kpc when the temperature of the plasma is $\sim 10^4$ K. Galactic winds, therefore, can be a major source of \ovi even at temperatures where it should not be present considering only equilibrium ionization, and may even have a significant contribution to \ovi foreground near the Solar neighborhood \citep[see][]{Breitschwerdt1994}. 

The inclusion of NEI physics together with the self-radiation (\neisr run) is shown by the solid curves in figure \ref{fig:O-ionfracs}. Given that both the NEI physics and the excess radiation field try to keep ion fractions high for the intermediate and high ions, the addition of the two mechanisms work together to keep the ion fractions even higher compared to both the \piesr and \nei cases. Lower ions (such as \oiii), on the other hand, are mostly dependent on the radiation field than the NEI effects. The addition of NEI physics, therefore, does not change the low ion fractions much.

It is interesting to note that both the self-radiation and the NEI physics affect the \ovi to almost the same extent in this particular case.
The relative importance of these two phenomena to dictate the \ovi ion fraction for other wind parameters will be clear once we explore the parameters. We also note that we find a negligible amount of \oii (ion fraction $\lesssim 10^{-5}$) in our simulations due to a higher photo-ionization rate at low temperatures. We speculate that any observable \oi or \oii in the wind is probably not a part of the freely expanding wind and must originate from other sources, such as the entrainment of cold clouds from the ISM and interaction between these clouds and the wind \citep{Prochaska2011, Scarlata2015, Krumholz2017, Fielding2022}  

\begin{figure*}
    \centering
    \includegraphics[width=\textwidth]{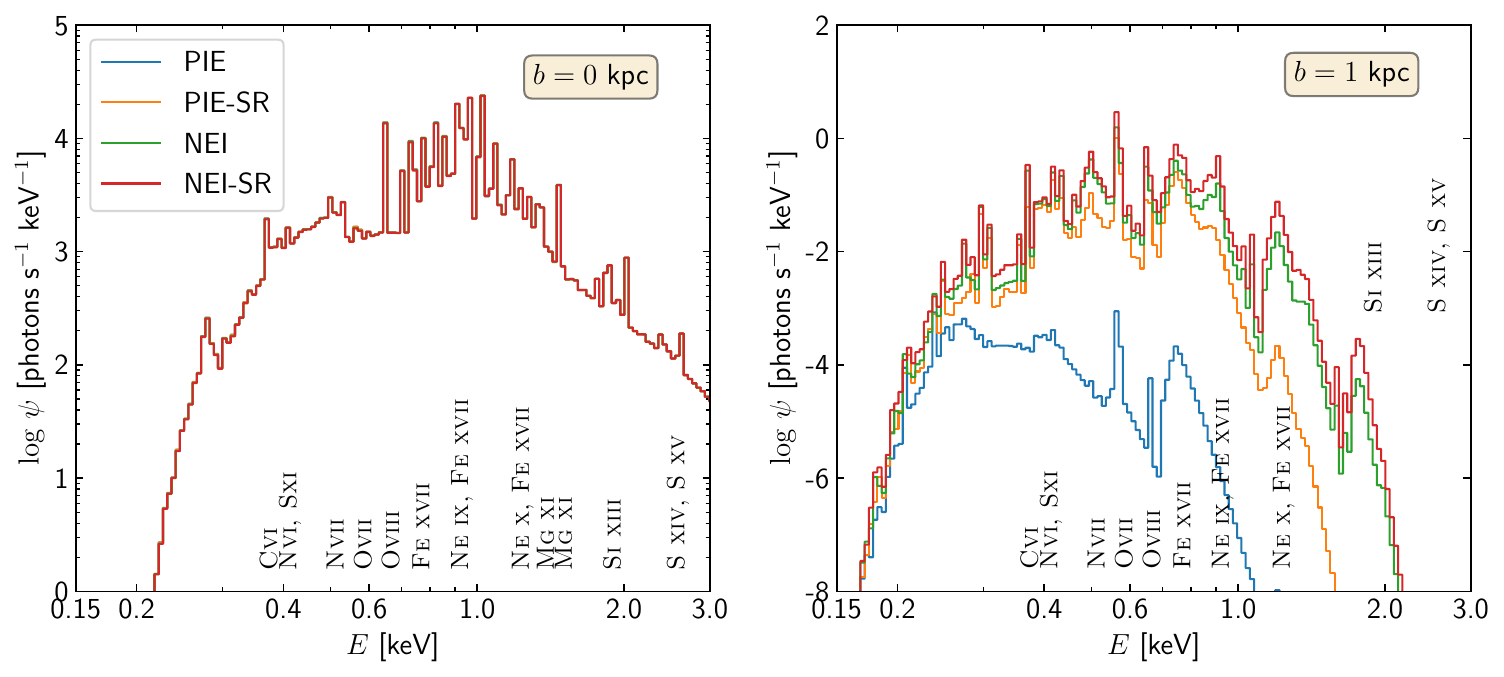}
    \caption{Observable emission spectra at different impact parameters, $b = 0$ (left panel) and $1$ kpc (right panel). The spectra is considered to be absorbed by the Galactic ISM at temperature $8,000$ K with column density $2\times 10^{21}$ cm$^2$. Non-equilibrium/self-radiation signatures are clearly observable for several lines, such as \ionr{O}{vii}, \ionr{O}{viii}, \ionr{Fe}{xvii} etc and corresponding free-bound continuum.}
    \label{fig:spectra-chandra}
    \vspace{0.5cm}
\end{figure*}
%
%
%
\subsection{Column densities}
\label{subsec:ion-columns}
We plot the column density of Oxygen ions for the different runs in figure \ref{fig:colden-oxygen}. Since in actual observations we often probe a projected density, we plot the Oxygen column density ($N_{\rm ion}$) as a function of the impact parameter, $b$, from the center, i.e. 
\begin{equation}
     N_{\rm ion} (b) = 2\times \int_{r=b}^{R_{\rm box}} n_{\rm ion}(r) \frac{dr}{\sqrt{1-b^2/r^2}}
     \label{eq:column-den}
\end{equation}
where, $n_{\rm ion}$ is the ion density and $R_{\rm box} = 10$ kpc, is the simulation box size. The overall decline in column density with radius is due to the decline in matter density. The sharp drop of ion column densities at $\gtrsim 7$ kpc is an artifact due to the limited simulation box size given that the length of the above integration is $\propto \sqrt{R_{\rm box}^2-b^2}/R_{\rm box}$. 

The \oiii and \oiv column densities at $b\approx 0$ (down-the-barrel) are suppressed by a factor of $5-10$ due to the combined effect of the NEI and self-radiation. The \ovi column density does not change much within $\sim 1$ kpc since it mostly follows temperature within this radius. The column density, however, increases by many orders of magnitude once either the self-radiation or the non-equilibrium physics is turned on. As a result, we expect to find \ovi at $r\gtrsim 2$ kpc, where it should not be present for equilibrium ionization (due to only stellar and HM12 radiation). A similar effect is also noticed in \oviii column density at $\gtrsim 2$ kpc where is the \oviii column is $\sim 3\times 10^{15} \pcm$ well within the observable limit. 

Figure 5 (see also Figure \ref{fig:colden_O_comp}) shows that the
various oxygen ions contributing to the total oxygen column density is 
determined by the non-equilibrium effects (for the high ions, 
\ovii and \oviii) and the 
radiation field  
(for the intermediate ions \oiii, \ovi etc). This has 
important implications 
for interpretation 
of observed wind ion emission and absorption spectra.
The figures show that the main oxygen stage is not \oiii (IP $=47.9$)
but is rather dominated by \ionr{O}{iv} (IP $= 75$ eV) and \ionr{O}{v} (IP $= 110$ eV). Conversion of just the \oiii to a total hydrogen column, assuming an oxygen abundance but without further "ionzation correction"  would lead to a severe underestimate of the mass and momentum outflow rates (c.f. \cite{Heckman2015}).
It is also possible that the observed \ovi emission around star-forming galaxies \citep[such as][]{Hayes2016}, may not be in collisional equilibrium but rather in photo-equilibrium. Given that the radiation field can be $\sim 10^4$ times higher than the UV background at $E \gtrsim 100$ eV (for SFR $= 10 \mpy$), we estimate that the critical density for photo-ionization of \ovi is $n_{\rm crit, ovi} \approx 0.5 \pcc$ at $T=3\times 10^5$ K. This density is very close to the estimated density of the \ovi emitting region in the starburst galaxy SDSS J115630.63+500822.1 \citep[SFR $= 20 \mpy$;][]{Hayes2016}.

\subsection{Emission spectra}
\label{subsec:emis-spec}
One key aspect of NEI physics is its effect on the observable X-ray emission. As we discussed above, higher ions such as \oviii are significantly over-ionized compared to the plasma temperature. This may produce emission line signatures for X-ray lines that are brighter than the equilibrium values, whereas, the continuum emission may simply follow the electron temperature. 

Figure \ref{fig:spectra-chandra} shows observable emission spectra for different simulations at two impact parameters, $b = 0$ and $1$ kpc. 
The emission spectra ($\ergpspcmsq$ sr $^{-1}$ Hz$^{-1}$) as a function of the impact parameter is calculated assuming optically thin emission as
\begin{equation}
    \psi_\nu'(b) =  2\times \int_{r=b}^{R_{\rm box}} \frac{\varepsilon_\nu(r)}{4\pi} \frac{dr}{\sqrt{1-b^2/r^2}}
\end{equation}
where, $\varepsilon_\nu(r)$ is the volume emissivity ($\ergpspcmsq$ Hz$^{-1}$) of the plasma at any given radial shell. We use \cloudy-17.01  for calculating non-equilibrium emissivity at each spatial grid for a given density, temperature, and non-equilibrium ion fractions.
We use a total 3780 frequency bins from $0.041$ eV to $12.5$ keV at a frequency resolution $\nu/\Delta \nu \approx 300$.
Once the emission spectra, $\psi_\nu'(b)$, is obtained we allow it to be absorbed by the ISM of our Galaxy, assumed to have gas at temperature $8,000$ K and column density $N_H = 2\times 10^{21}$ cm$^2$.
The observable spectra is, therefore, $\psi_\nu(b) = \psi_\nu'(b)\, \exp(-N_H \sigma_\nu)$, where $\sigma_\nu$ is the total photo-ionization cross-section of the absorbing gas, again obtained from \cloudy-17.01. The resulting spectra is then coarse grained at a frequency resolution of $\nu/\Delta\nu = 50$, corresponding to typical resolution of \textsc{Chandra} spectrometers at energy $\sim$ keV. Although the actual observations may have even better energy resolution, we use a coarser resolution for the sake of a better visibility in the plot. 
While our simulated spectra are in units of $\ergpspcmsq$ Hz$^{-1}$ sr$^{-1}$, the observable spectra are often reported  in photons$\psec$ keV$^{-1}$. We, therefore, convert the CGS flux to photon counts by assuming the
effective area and field of view to be $300$ cm$^2$ and $16'$. Note that the assumed Chandra specifications are only to provide a rough estimation of the total counts. Exact counts may depend on several factors including the variation of the  effective area, energy resolution and, and the efficiency of the spectrometer with energy. Additionally, contribution from the local bubble and cosmic X-ray background are not included in our simplistic spectrum calculation. Detailed modeling of the observable spectra through any instrument is out of the scope of this paper. 

\begin{figure*}
    \centering
    \includegraphics[width=0.9\textwidth]{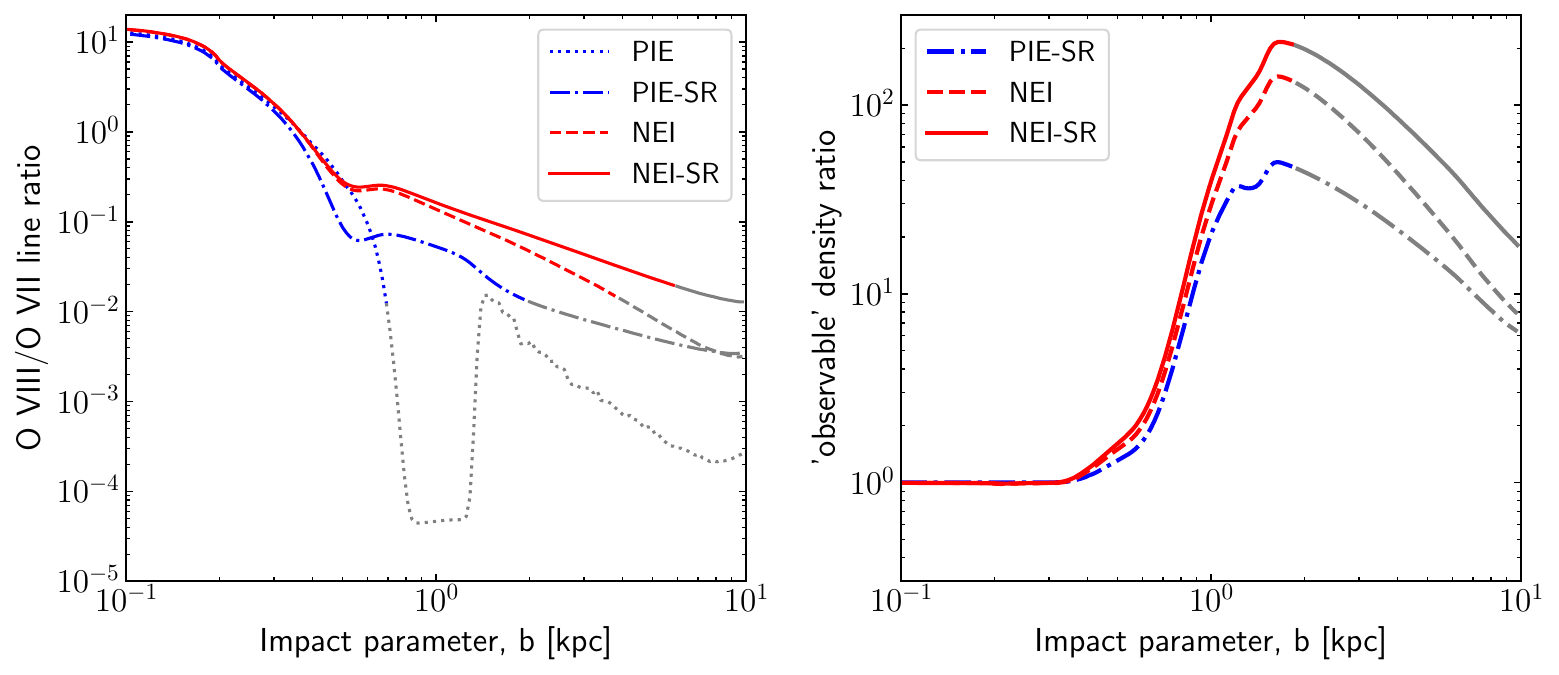}
    \caption{Left: Ratio of line intensities of two oxygen lines, \ionr{O}{viii} ($18.97 \angs$) and \ionr{O}{vii} ($22.1012\angs + 21.6020\angs + 21.8070\angs$). Different line-styles represent different simulations. The gray part of each line represents the region where the \ionr{O}{vii} ($22.1012\angs + 21.6020\angs + 21.8070\angs$) intensity falls below $10^{-5}$ of the central ($b=0$) value. This low-intensity region is not expected to be observed. Right: `Observable density' ratio (equation \ref{eq:observable-density}) of \piesr, \nei, and \neisr with respect to the \pie case. The gray parts of the lines represent the region when the `observable density' in \pie case is lower than $10^{-5} \times$ its central value.}
    \label{fig:line-ratio}
    \vspace{0.5cm}
\end{figure*}

The `observable spectra' in figure \ref{fig:spectra-chandra} also show the approximate position of several emission lines. The figure shows that the observable spectra for the different runs with varying physics are not very different at $b =0$ (down-the-barrel). This is expected since most of the emission at this impact parameter comes from the central star-forming region where the plasma is in collisional equilibrium. The spectra for the different runs at $b = 1$ kpc, however, show much deviation from each other, especially at the energy range $0.3-1.5$ keV. 
We note several dominating line emissions, such as \ionr{O}{vii}, \ionr{O}{viii} and \ionr{Fe}{xvii} in this energy range, that vary more than a few orders of magnitude among different runs. For example, the \ionr{O}{vii} line at $0.57$ eV increases from $0.001$ in case of \pie to $\sim 1 \photpskeV$ in \piesr. This is due to the additional photo-ionization from the excess hard radiation produced by the central hot gas.
The line rises to $2\: \photpskeV$ for the \nei case due to the delayed recombination of \ionr{O}{vii}. Addition of both the NEI network and self-radiation increases line intensity to $3\: \photpskeV$ due to increased \ionr{O}{vii} column density as explained in section \ref{subsec:ion-fracs} and \ref{subsec:ion-columns}. This same effect can also be noticed in many other lines as shown in the figure. Note that although the non-equilibrium plasma models predict higher line emissions in the X-ray bands, the total un-absorbed emissivity (integrated over the whole frequency range) is slightly smaller for these models. The higher X-ray emissivity in non-equilibrium models, therefore, are consistent with the known suppression of overall cooling efficiency in recombining plasma \citep{Gnat2007, Gnat2017}.

\subsection{Observable temperature and density}
This excess line emission has direct effects toward the determination of ion temperatures in the plasma. Figure \ref{fig:line-ratio} shows the emission line ratios between \ionr{O}{viii} ($18.97 \angs$) and \ionr{O}{vii} ($22.1012\angs + 21.6020\angs + 21.8070\angs$) lines, an often used indicator for the temperature in X-ray emitting plasma \citep[e.g.][]{Henley2010, Henley2012, Miller2015, Sarkar2017}. A higher value of this ratio indicates a higher temperature for the plasma. The figure shows that the line ratio drops quickly with the impact parameter for the \pie case, implicating a temperature drop. However, \piesr does not show this sudden drop and the line ratio decreases slowly. The \nei and \neisr cases show a similar behavior as in \piesr but the line ratio remains relatively high for the most part of the wind. For example, if we consider the line ratio at $\sim 1$ kpc, \pie represents a temperature $\sim 3\times 10^5$ K corresponding to ratio of $\sim 10^{-4}$, whereas, the \nei and \neisr values ($\sim 0.3$) represent a temperature of $\sim 4\times 10^6$ K. Therefore, fitting the X-ray spectra using only CIE or PIE would lead to an overestimate of the actual temperature in the wind. Although the temperature of the x-ray emitting wind is usually estimated by fitting the whole x-ray spectra i.e. simultaneously fitting the lines and the continuum \citep{Strickland2002, Lopez2020}, we only use the two lines as an indicator of the the temperature for simplicity.

A similar effect can also be noticed for the observed density. We plot the `observable density' ratio for different cases compared to the \pie case in the right panel of figure \ref{fig:line-ratio}. We define the `observable density', $\tilde{n}$, assuming that the total X-ray emission is $\propto \tilde{n}^2$ at any given impact parameter i.e.,  
\begin{equation}
    \tilde{n}^2 = \int_{0.2 {\rm keV}}^{3 {\rm keV}} \psi(E)\, dE
    \label{eq:observable-density}
\end{equation}
where, $\psi(E)$ is the photon count in $\photpskeV$ from figure \ref{fig:spectra-chandra}. Note that this is not the actual density of the plasma rather an arbitrarily scaled version of the actual value. While the actual density values depends on the details of the X-ray spectral fitting, the ratio of $\tilde{n}$ between different cases should roughly represent the ratio of the actual densities in these cases. Figure \ref{fig:line-ratio} shows that with increasing non-equilibrium and self-radiation effects, the observable density can be higher by a factor of a $\sim 10$ (at $\sim 1$ kpc) to two orders of magnitude ($\sim 2$ kpc). This overestimation is due to the excess line emission from the over-ionized X-ray lines.

We, therefore, speculate that X-ray observations of galactic winds  may overestimate the temperature and mass carried out by the wind \citep[e.g.][]{Strickland2000, Lopez2020} unless accounting for the non-equilibrium and the self-radiation effects. Detailed analysis of the existing (such as \textsc{Chandra}, \textsc{Suzaku}) or expected X-ray spectra from the upcoming missions (such as \textsc{xrism} and \textsc{athena}) will enable us to study these effects.

\section{Discussion}
Our results show the necessity to consider the non-equilibrium ionization (NEI) and self-radiation physics in galactic winds. We now discuss several implications.  

\subsection{Multi-phase interaction}
\label{subsec:multiphase-interaction}
An important aspect of wind thermodynamics and observable emission is the multiphase interaction with the cold clouds. Warm/Cold clouds from the ISM are often entrained by the wind and are observed to be accelerated to $\sim 100 \kmps$ \citep{Heckman1990, Shopbell1998, Martin1999}. Observations of the warm/cold clouds in the galactic winds are based on several emission/absorption lines, such as H$\alpha$, [\ionr{N}{ii}], [\ionr{O}{ii}], \ionr{Na}{I} etc. \citep{Heckman1990, Cecil2001, Strickland2004a, Martin2005, Heckman2015, Steidel2010}, especially in galaxies where hot X-ray emitting gas is not observable. A proper estimation of the total mass flow rate in these observations is crucial since it gives first-hand information about the effectiveness of feedback mechanism in that galaxy. Such estimates are, however, dependent on several assumptions including on the ionization state. As we have pointed out in the current paper, the ionization state of the wind depends heavily on the NEI effects and the radiation field. The excess hard-radiation field is also expected to affect the ionization of the warm/cold gas entrained by the wind, an effect that needs further investigation. 

Interaction of wind and cold gas, and its observable signatures are important parts of the above question and is currently under active investigation \citep{Kwak2010, Kwak2015, Armillotta2017, Gronke2018, Fielding2020, Kanjilal2021}. Although most of the current efforts model the thermodynamics of the interaction, only a handful consider the NEI of ions to model the observable emission/absorption lines \citep{Kwak2010, Ji2019}. Excess radiation field in the galactic wind can also affect ionization states in the interaction layer and increase/decrease its efficiency through additional heating/cooling. Our discovery of the excess hard radiation implies the need for more complicated considerations when modeling the emission/absorption signatures of warm/cold gas in the wind.

While the effect of the wind on the warm/cold clouds is acknowledged, we often neglect the effect of the cloud on the wind itself. Recent simulations by \cite{Schneider2020} show that the mixing of warm/cold clouds with hot gas significantly changes the thermodynamic profiles of the wind. The mixing of the cold gas with the wind increases the density of the wind and makes the density profile flatter than $\rho \propto r^{-2}$. The interaction also slows down the dynamical cooling of the wind due to the creation of bow shocks at the head of the clouds \citep{Cooper2008, Vijayan2018} and thereby converting a fraction of the overall kinetic energy to thermal energy. Description of the wind including such interactions has recently been studied to some extent \citep{Fielding2022, Nguyen2021}. Given that the dynamical cooling time increases due to wind-cloud interaction, we speculate from sec \ref{sec:intro} that the NEI effects on the wind would be suppressed. On the other hand, the excess hard radiation may increase due to the increased density of the wind which can ionize the wind further. The exact amount of suppression of NEI effects and increase in self-radiation depends on the profile of the wind. 

%
%
%

\subsection{Radiation field into the CGM}
\label{subsec:radfield-cgm}
The excess X-ray photons can also be an important source of hard radiation in the CGM. The angle averaged specific intensity at any $r \geq R$ can be roughly be estimated from equation \ref{eq:N_XC}. A simple correction due to the shape of the spectrum (comparing the estimated value in equation \ref{eq:N_XC} and fig \ref{fig:spectra-nphots} at $1$ keV) produces
\begin{eqnarray}
N_X(r) &\approx& 5\times 10^4\, \frac{\beta^3\:  \Lambda_{-23}}{\epsilon\,R_{100 {\rm pc}}^3\, \bar{E}_{\rm keV}} \left(\frac{SFR}{\mpy}\right)^2\, \left(\frac{R}{r}\right)^2 \nonumber \\
  && \quad\quad\quad \photps \pcmsq \mbox{ sr}^{-1} 
\end{eqnarray}
We can now estimate the region over which the hard-radiation intensity from the wind will be dominant over the HM12 background intensity ($N_{X,{\rm hm12}}$), i.e. setting $N_X(r) \gtrsim N_{X,{\rm hm12}}$. This produces a critical radius beyond which the hard-radiation from the wind becomes sub-dominant. This radius is given as 
\begin{equation}
    r_{\rm crit} \approx 7 \mbox{  kpc  } \left(\frac{SFR}{\mpy}\right) \times \left[ \frac{\beta^3 \Lambda_{-23}}{\epsilon\,R_{100 {\rm pc}}\, \bar{E}_{\rm keV}} \right]^{1/2} \,,
    \label{eq:rcrit-cgm}
\end{equation}
where we have used $N_{X,{\rm hm12}} = 10 \photps \pcmsq$ sr$^{-1}$ from fig \ref{fig:spectra-nphots}. For the wind parameters used in the current paper, we estimate $r_{\rm crit} \approx 40$ kpc. Although the above equation suggests that increasing the mass loading factor ($\beta$) would increase $r_{\rm crit}$, it may not be the case for an arbitrarily large $\beta$. For very high $\beta$ values, the central temperature of the wind would decrease and hence decreasing the soft X-ray emission function, $\Lambda_{-23}$.  Additionally, a very high value of $\beta$ may also stop launching supersonic winds due to very high radiative cooling inside the star forming region \citep{Thompson2016}. As mentioned earlier, equation \ref{eq:rcrit-cgm} only provides an order of magnitude estimate.

\subsection{Non-spherical wind}
\label{subsec:non-spheri-wind}
While the classic description of spherically symmetric winds often provides a first-order assessment of the wind thermodynamics, realistic winds are often non-spherical and show bi-conical geometry about the galactic plane \citep{Lopez2020}. A non-spherical wind does not necessarily change the overall steady-state profile of the wind if the solid angle of the wind does not change with radius (since the divergence terms in equation \ref{eq:mass-eq} do not change). In such a case, the thermodynamic quantities are only normalized to maintain the same flow rate. The profile, however, changes if the solid angle of the wind changes with radius. It has been recently shown by \cite{Nguyen2021} that the thermodynamic profiles become much flatter than the spherically symmetric wind and are dependent on the profile of the solid angle. While \cite{Nguyen2021} show its applicability to the M82 galaxy, a general application to galaxies would require further investigation of how outflows are collimated and if there is any apriori way to predicting the wind-opening angle. 

\subsection{Effects of ISM absorption}
\label{subsec:ism-abs}
In our  
spherical wind model, we find that the wind can completely clear out the ambient ISM and establish a steady state wind. 
However, realistic star-forming regions may 
only be partially porous to the wind. This is especially true for 
weaker winds 
that cannot clear out the ambient ISM completely. While hot gas can still percolate out via the porous medium and set up a steady state wind, the radiation from stars and hot gas may be at least partially absorbed. The amount of absorption depends on the ISM column density and the frequency of the photons. The fraction of escaped radiation 
at $r>R$ can be be estimated as 
\begin{equation}
    \frac{J'_\nu(r)}{J_\nu(r)} \sim \left(1-f_{\rm cov}\right) + \frac{R'(\tau_\nu=1)}{R}
    \label{eq:ism-abs}
\end{equation}
where, $R'$ is the skin-depth of the star-forming region below which a particular frequency is not observable i.e., optical depth, $\tau_\nu = 1$. This expression holds well for the frequencies that are expected to be highly absorbed. For such absorbed frequencies, the visible volume is $V' \sim 2\pi R^2 R'$ and the fraction of escaped radiation is then $\sim V'/(4\pi/3 R^3) \sim R'/R$. 

The skin-depth can be estimated using $\sigma_\nu n_H  R'= 1$, where $\sigma_\nu$ is the photo-ionization cross-section 
at a given frequency $\nu$. Now if the total ISM column density is $N_H$, then $n_H = N_H/2 R $, therefore, the skin-depth is given as $R'/R = 2/(\sigma_\nu N_H)$. For example, if we assume $N_H = 2\times 10^{21} \pcmsq$ for a neutral ISM, then $R'/R \sim 10^{-4}$, $10^{-2}$ for energies $E = 13.6, 100$ eV. That means the \ionr{H}{i} ionizing photons will be mostly absorbed while only a small fraction of the soft x-ray photons escape. However, note that this is highly dependent on the assumed column density of ISM that is blocking the radiation.

We also simulated a slightly modified version of \neisr to see the effect of ISM absorption inside the star-forming region. This simulation, in addition to the \neisr physics, considers the opacity at every cell of the star forming region is due to a plasma at $T = 8,000$ K and hydrogen density of $n_H = 1.62 \pcc$, corresponding to a total hydrogen column density of $2\times 10^{21} \pcmsq$ and $f_{\rm cov}=1$. The results for this simulation are shown in figures \ref{fig:spectra-nphots-ism} and \ref{fig:O-ionfracs-ism}. Consistent with our estimates, we see that the radiation field is now indeed $\sim 10^{-4},\: 10^{-2}$ times lower at $13.6,\:100$ eV. Since the ionization field is now significantly reduced, the ion fractions now are similar to the \nei case only (Figure \ref{fig:O-ionfracs-ism}). The wind ion fractions, therefore, are sensitive to the total absorbing column density. However, at such high ISM absorption, the radiation field is mostly dependent on the covering fraction, $f_{\rm cov}$. 
For high star-formation rates 
the star forming region can be highly over-pressured and can often clear out a big enough channel through which both the hot plasma and radiation can escape easily \citep{Sarkar2015a, Fielding2018}. 
While ISM absorption could reduce the radiation emanating from weakly
star-forming regions that have not cleared out large channels, such absorptions are likely much less
important for high enough star-formation rates driving energetic winds.

%
\section{summary}
\label{sec:summary}
Galactic winds are central feature of feedback regulated galaxy evolution. We perform numerical studies of spherically symmetric galactic winds from star-forming galaxies including the effect of non-equilibrium ionization of the plasma and the radiation from the young stars, cosmic UV background, and importantly, the self radiation of the plasma itself. Our findings of the wind can be summarized as follows:
\begin{itemize}
    \item Galactic winds can be a major source of the galactic soft X-ray radiation ($100$ eV $\lesssim E \lesssim 1$ keV). This is due to the dense and hot plasma produced by the energy deposition of supernovae remnants within a small region. The radiation can be about a few orders of magnitude higher compared to the \citep{Haardt2012} UV background within $10$ kpc. The excess radiation can even surpass the background UV radiation within the central $\sim 50$ kpc depending on the wind parameters (such as the mass loading factor, thermalization coefficient, and star formation rate). The radiation at $E \lesssim 100$ eV is mostly dominated by the radiation from the young O/B stars.
    
    \item The presence of the excess soft X-ray radiation is sufficient to ionize the wind itself. In addition to being able to ionize the low ions (such as \ionr{O}{iii}), the excess radiation can also keep X-ray emitting ions (such as \ionr{O}{vi} and \ionr{O}{vii}) ionized to a few orders of magnitude higher compared to their collisional ionization level. 
    
    \item Even without the excess radiation, X-ray emitting ions are often in a non-equilibrium ionization state compared to their temperature due to longer effective recombination times. The galactic winds are often over-ionized. The observable metal column densities for high ions (such as \ionr{O}{viii}) at a given impact parameter of the wind can vary by almost a few orders of magnitude due to the non-equilibrium effects. The lower ions (such as \ionr{O}{iii}) on the other hand are more affected by the presence of the excess radiation field than the non-equilibrium ionization physics.
    
    \item Our synthetic X-ray spectra of the wind predict several line emissions to be highly affected by the presence of the non-equilibrium and radiation field which may lead to inaccurate density or temperature estimations for the wind if these effects are not accounted for. 
\end{itemize}

In conclusion, we have shown that the effects of non-equilibrium ionization and self-radiation are of crucial importance in understanding galactic wind observations. Challenges, however, remain to include a few other thermodynamic effects before our galactic wind model can be directly used to quantitatively interpret the observations. At the same time, we should also prepare better spectrum analysis tools to correct for the presence of non-equilibrium effects in the current (such as \textsc{chandra, XMM-newton}) as well as the future generation of X-ray telescopes, such as \textsc{xrism} and  \textsc{athena}.

%
%
%
%
\begin{acknowledgments}
This work was supported by the German Science Foundation via DFG/DIP grant STE 1869/2-1 GE625/17-1 at Tel-Aviv University and the Israeli Science Foundation (ISF grant no. 2190/20), and by the Center for Computational Astrophysics (CCA) of the Flatiron Institute and the Mathematics and Physical Sciences (MPS) division of the Simons Foundation, USA.
\end{acknowledgments}


\software{\cloudy \citep{Cloudy2017}, \pluto \citep{Mignone2007}, \textsc{matplotlib} \citep{matplotlib}. }
 
\bibliography{library}
\bibliographystyle{aasjournal}

\appendix
\section{Equivalent stellar emissivity}
\label{app-sec:equiv-emissivity}
Stellar radiation from stars in a star-forming region, $R$, can be approximated to an equivalent emissivity per unit volume. This bypasses the requirement to consider the emission arising from individual stars. The equivalent emissivity can be found once we assume that the stars are distributed randomly across the whole star-forming region such that one can define the number density  of stars to be 
\begin{equation}
    n_\star = \frac{3 N_\star}{4\pi R^3}
\end{equation}
where, $N_\star$ is the total number of stars, each with radius $R_\star$ and surface brightness $B_\star$. Therefore, the specific intensity at any given ray direction, $\mu$ ($= \cos\theta$; $\theta$ is the angle between the ray direction and the radial direction), at the edge of the star-forming region is $\psi_\star(\mu) = A_\star \times B_\star  = 3 \mu R_\star^2 B_\star N_\star/(2 R^2)$. Here we used the fact that the radiation to this direction only comes from a cylinder of unit surface area but length equal to the chord corresponding to the ray direction so that the total stellar surface area within this unit cylinder is $A_\star = 2\mu R \times n_\star \times \pi R_\star^2$. Now, the total luminosity of each of the stars is $l_\star = 4\pi R_\star^2 \times \pi B_\star$. Therefore, 
\begin{equation}
    \psi_\star (\mu) = \frac{3 l_\star N_\star}{ 4\pi R^2} \frac{\mu}{2\pi} = \frac{3 L_\star}{ 4\pi R^2} \frac{\mu}{2\pi}
    \label{app-eq:stellar-psi}
\end{equation}
where, $L_\star = l_\star N_\star$ is the total luminosity of the stars. 

Now if the equivalent emissivity ($\varepsilon_\star$) is assumed to be uniform throughout the star-forming region, the radiation at $r=R$ for the same unit cylinder is $\psi_\star(\mu) = 2 \mu R \times \varepsilon_\star/(4\pi)$. A quick look at Eq \ref{app-eq:stellar-psi} then produces the equivalent emissivity to be
\begin{equation}
    \varepsilon_\star = \frac{3 L_\star}{4\pi R^3}\,.
\end{equation}
This is the same emissivity as if the total stellar radiation is assumed to be a scalar quantity. Note that once the effective emissivity is known, equation \ref{app-eq:stellar-psi} can also be used to get the radiation field at any external point for radiation generated by the hot plasma.

\section{Ionization parameter}
\begin{figure}
    \centering
    \includegraphics[width=0.45\textwidth]{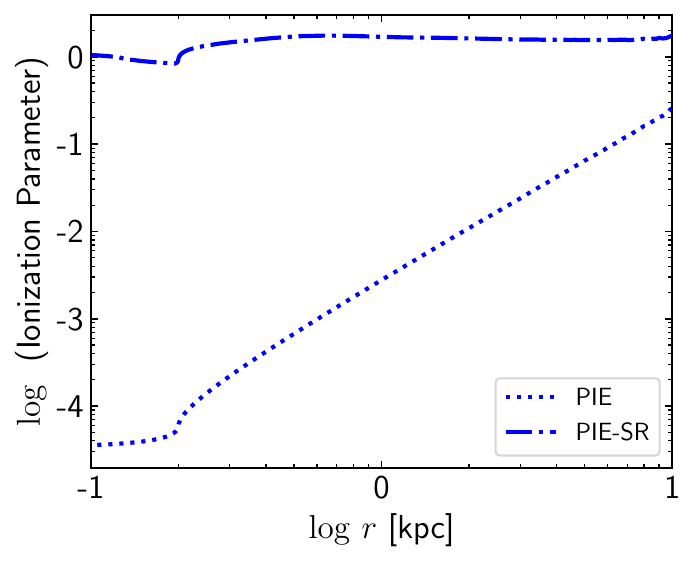}
    \caption{Ionization parameter for \ionr{O}{vi} (ionization potential $= 138$ eV) at steady state for \pie (dotted line) and for \piesr runs (dash-dotted line). The sharp rise in ionization parameter in \pie case happens due to the sharp decrease of the overall density. The ionization parameter remains roughly constant for the \piesr case since both the density and radiation field decreases as $\sim 1/r^2$.}
    \label{app-fig:ioniz-param}
\end{figure}

The presence of radiation field strongly affects the ionization fraction of lower as well the intermediate ions, such as \ovi. A comprehensive way to understand this is to look at the ionization parameters. We define the ionization parameter for any ion $i$ of belonging to an element $k$ as
\begin{equation}
    \chi_{k,i} = \frac{4\pi}{n_{\rm k} c} \int_{\nu\geq \nu_{0,k,i}} \frac{J_\nu}{h_p \nu} d\nu  
\end{equation}
where, $h_p \nu_{0,k,i}$ is the ionization potential of the ion ($k$,$i$) and $n_k$ is the volume density of the element, $k$. 
Figure \ref{app-fig:ioniz-param} shows the ionization parameters of \ovi for the cases with and without self-radiation. Although the ionization parameter of \ovi for the \piesr case remain close to unity throughout, the actual ion-fraction will be determined by the strength of the radiation field and the collisional ionization rate coefficients ($\xi_{k,i}$) at that location. 

\section{ISM absorption}
\begin{figure*}
    \centering
    \includegraphics[width=0.75\textwidth]{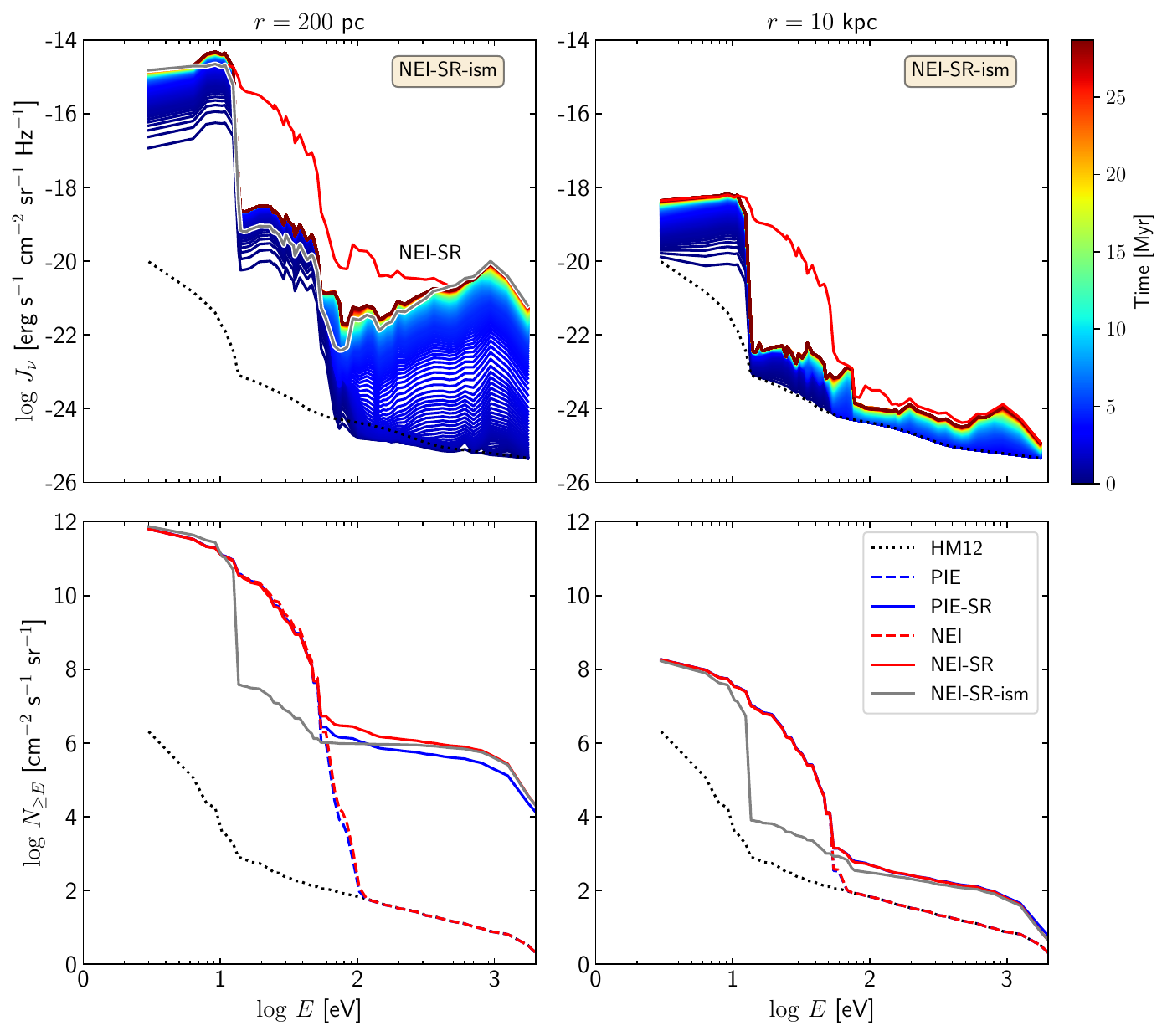}
    \caption{Radiation spectra inside the wind (same as figure \ref{fig:spectra-nphots} but with ISM absorption. \textit{Top panel}: Evolution of the spectrum. The solid red line represents the un-absorbed spectra at $t=28$ Myr. The predicted ISM spectra (\ref{eq:ism-abs}) is shown in gray solid line. The reduction of the radiation field is due to absorption from ISM ( corresponding to $N_H = 2\times 10^{21} \pcmsq$, $T = 8,000$ K) uniformly distributed within the star-forming region. \textit{Bottom panel}: Total number of photons with energy $> E$. The solid gray line shows the final spectra (at $t = 28$ Myr) for the un-absorbed case.}
    \label{fig:spectra-nphots-ism}
\end{figure*}

\begin{figure}
    \centering
    \includegraphics[width=0.43\textwidth]{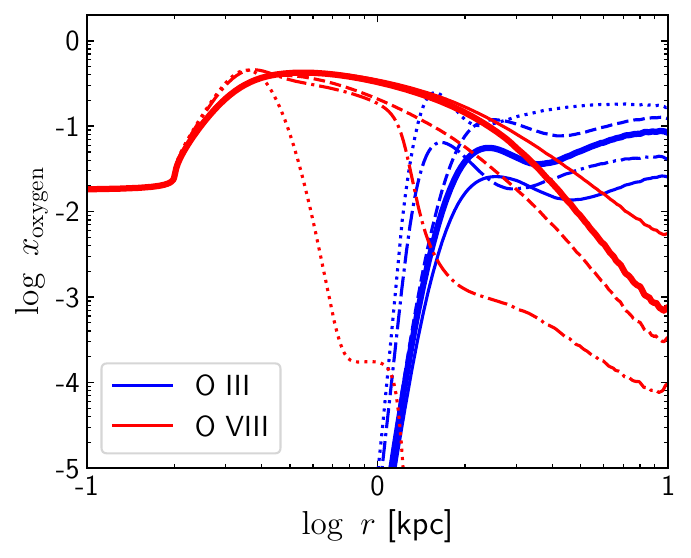}
    \caption{Ion fractions of \ionr{O}{iii} and \ionr{O}{viii} in the wind (same as figure \ref{fig:O-ionfracs}) but shows results for simulations containing ISM absorption (thick solid lines). Different line styles represent different simulations -  \textit{dotted}: \pie, \textit{dashed}: \nei, \textit{dash-dotted}: \piesr, and \textit{solid}: \neisr.}
    \label{fig:O-ionfracs-ism}
\end{figure}

\begin{figure}
    \centering
    \includegraphics[width=0.43\textwidth]{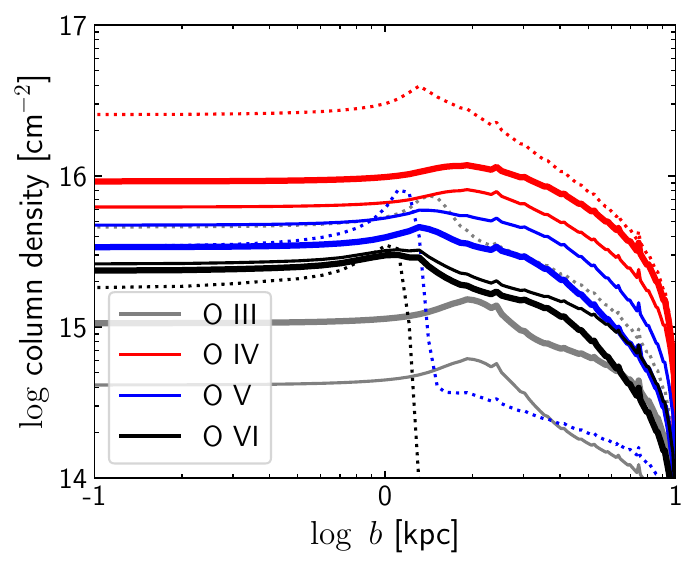}
    \caption{Column densities for `warm' ions showing that the \ionr{O}{iv} (IP $=75$ eV) and \ionr{O}{v} (IP $=110$ eV) dominate the total oxygen outflowing mass. Different line styles are - \textit{dotted}: \pie, \textit{thin-solid}: \neisr, \textit{thick-solid}: \neisr with ISM absorption. }
    \label{fig:colden_O_comp}
\end{figure}
\label{lastpage}
\end{document}